\documentclass[useAMS,usenatbib,usegraphicx,dvips]{mn2e}

\newcommand{\et}{\sl et al. \rm}

\title[AGN UV peak : blurred reflection ?]{The UV peak in Active Galactic Nuclei  : a false continuum from blurred reflection ?}
  
\author[Lawrence]{
 A.Lawrence$^{1}$ \\
 Revised version : March 2nd 2012 \\
$^{1}$Institute for Astronomy, SUPA (Scottish Universities Physics Alliance), University of Edinburgh, \\ 
Royal Observatory, Blackford Hill, Edinburgh EH9 3HJ, UK \\
}
\begin{document}

\date{Accepted 2012 March 8th. Received 2012 March 2nd ; in original form 2011 July 18}

\pagerange{\pageref{firstpage}--\pageref{lastpage}} \pubyear{2012}

\maketitle

\label{firstpage}


\begin{abstract}
I summarise and analyse key problems with observations of the UV bump in Active Galactic Nuclei (AGN), and especially the accretion disc interpretation - the temperature problem, the ionisation problem, the timescale problem, and the co-ordination problem - and suggest that all these problems can be solved if, in addition to the accretion disc, there is a population of cold, thick clouds at approximately 30$R_S$ which reprocess the intrinsic continuum. Exploring cloud parameter space, I find that clouds with density $n\sim 10^{12}$cm$^{-3}$ and column $N_H >4\times 10^{24}$cm$^{-2}$ reflect most of the intrinsic continuum, but convert a substantial fraction of the EUV luminosity into lines, dominated by Ly$\beta$ and HeII Ly$\alpha$. When velocity-blurred, this makes a false continuum peak at $\sim$1100\AA\ which fits the observed SED well, but turns back up in the FUV to make a hard EUV SED, as required by ionisation models. I argue that the observed UV variability is dominated by this component of fixed shape, possibly due to changes of covering factor. The amount of mass required is small, so it is not necessary to disrupt the disc, but only to make an unstable and inhomogeneous atmosphere. The proposed clouds may be related to those suggested by several X-ray phenomena (X-ray reflection components, high-velocity outflows, Compton thick partial covering) but are not the same, leading to a picture with a wide range of inhomogeneous structures at different radii.

\end{abstract}



\section{Introduction} \label{sec:intro}

The most prominent feature in the Spectral Energy Distribution (SED) of Active Galactic Nuclei (AGN) is the so-called ``Big Blue Bump'' (BBB).  This broad feature peaks in the UV \citep{Sanders1989, Elvis1994}, roughly consistent with the expectation of thermal emission from gas accreting onto a black hole mass with a mass of $10^{7-9} M_\odot$. The BBB is however broader than a single black body, as one would expect if accretion takes place in a disc with a range of temperatures at different radii \citep{Shields1978,Malkan1982}. In more detail, modellers have found it hard to reproduce the shape of the SED, especially near the UV peak, and through the soft X-ray region (e.g. \citealt{Blaes2001, HaroCorzo2007}).

Some observations provide generic support for the idea of a disc-like structure in the centre of AGN. (i) The strong Fe line and hard X-ray flattening seen in the X-ray spectra of AGN are often interpreted as being due to reflection by  ``cold'' material that covers a solid angle of $\sim 2\pi$ as seen by the X-ray source \citep{Pounds1990}.
 (ii) In a handful of quasars, optical spectra in polarised light, removing contaminating emission, show a broad dip possibly corresponding to the Balmer edge absorption expected from an accretion disc atmosphere \citep{Kishimoto2003}. Extending the polarised continuum into the near-IR reveals the classic long wavelength  $\nu^{1/3}$ spectrum expected from simple accretion disc models \citep{Kishimoto2008}.  (iii) For radio loud objects, and especially superluminal objects, where we can have some indication of the orientation of the sources, AGN show velocity width and and surface brightness correlations with viewing angle \citep{Brotherton1996, Rokaki2003, Jarvis2006}. (iv) At optical wavelengths, over a very broad range of luminosities AGN show colour-luminosity trends consistent with the generic expectation from multi-temperature black bodies \citep{Lawrence2005}.

It is notable that all the above successes concern optical or near-IR wavelength emission. In the UV however a number of observations disagree badly with at least the simplest accretion disc models. These issues can be characterised as the temperature problem, the ionisation problem, the timescale problem, the co-ordination problem, and the size problem. Some of these problems, along with other key issues such as polarisation predictions, have been discussed by \citet{Antonucci2002} and references therein. In the next section, I look at the current state of each of these key problems in turn. In the following section, I consider an idea that has often been seen as a rival to accretion disc models - emission from dense clouds. In fact, a {\em combination} of accretion disc and  dense inner clouds may be exactly what is needed to solve the problems of accretion disc models, as large velocity blurring can produce a localised false peak in the SED. Finally, I briefly discuss implications of this idea, and how it can be tested.

\section{Problems with the Big Blue Bump}  \label{sec:probs}

Here I summarise a number of problems with our current understanding of the Big Blue Bump.

\subsection{The temperature problem} 

AGN seem to be cooler than they ought to be. The SEDs of AGN seem to show a universal near-UV shape, reaching a maximum in $\nu S_\nu$ at a wavelength of around 1100\AA , more or regardless of luminosity or redshift \citep{Zheng1997, Telfer2002, Scott2004, Shang2005, Binette2005}.  Such a peak suggests a characteristic temperature of T$\sim$ 30,000K, and indeed this is similar to the inner temperature of accretion disc fits ever since \citet{Malkan1982}. However, for any thermal model, the characteristic temperature should be roughly 

\[ T_{ch} = {\rm 95,000} L_E^{1/4} M_9^{-1/4} R_5^{-1/2} \]

\noindent where $L_E = L/L_{Edd}$ is the luminosity in units of the Eddington luminosity, $M_9$ is the mass of the central object in units of 10$^9 M_\odot$, and $R_5$ is radius in units of  $5R_S$, where $R_S$ is the Schwarzschild radius. Here $R_5$ is taken to be the radius of an optically thick spherical surface from which blackbody radiation at temperature $T$ emerges. Other estimates, such as the radiation-density equivalent temperature, will be similar, and other possible characteristic temperatures, such as the Compton heating temperature, will be hotter, and rather model dependent \citep{Ferland1988}. 
The maximum temperature of a simple accretion disc extending to the last stable orbit is substantially hotter than the naive temperature above. Discs in which one imposes the zero-stress condition at the inner boundary can be cooler in the middle and hotter towards the outer radii, but it is far from clear whether this applies to black hole accretion discs (see for example \citealt{Frank2002}, p.90). That detailed accretion disc models do not overcome this problem is clear for example from the attempts to fit the SED of 3C273 by \citet{Blaes2001}. Smaller black holes are hotter, but objects radiating at much less than the Eddington limit are cooler, so for local Seyferts, as opposed to luminous quasars, we might expect some diversity; but these also seem to show the universal knee.

\begin{figure*}
\centering
\includegraphics[width=0.5\textwidth,angle=-90]{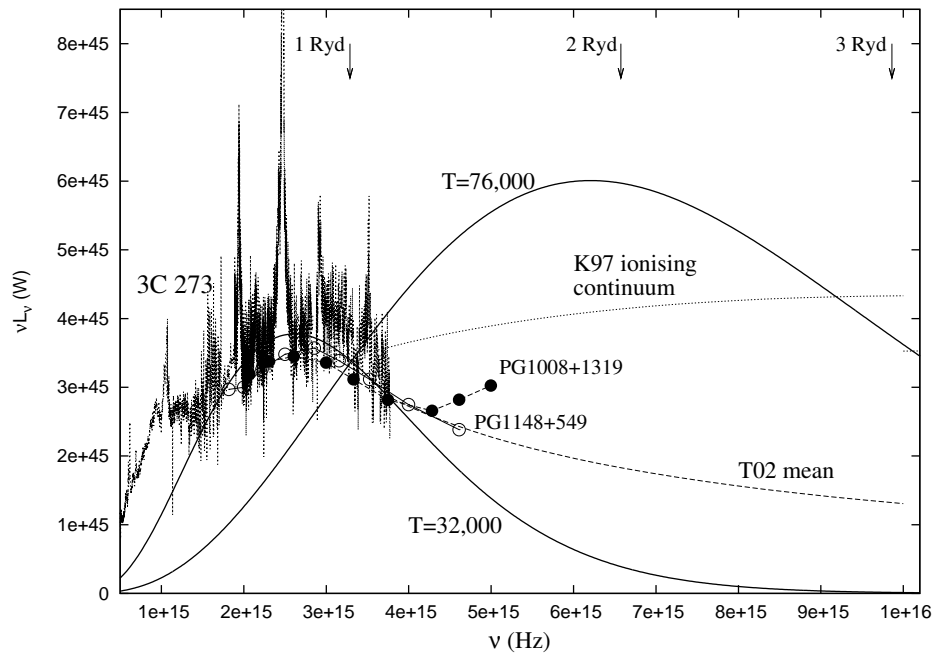}

\caption{\it\small Comparisons between various observed SEDs and simple theoretical curves. Note that both axes are linear, which helps to emphasise how strongly the SED is dominated by the NUV peak. The dotted lines show the SED of 3C~273 taken from \citet{Kriss1999}, kindly supplied by G.Kriss, using nearly simultaneous data from various instruments. The datapoints have been de-reddened assuming E(B-V)=0.032 (see \citealt{Kriss1999}). The circles show continuum from two quasars analysed by \citet{Binette2005}, estimated from their figures. The dashed line labelled ``T02'' shows the composite SED of \citet{Telfer2002}, represented as a power law with index -1.8. The blackbody curve with T=32,000K is chosen as an illustration of how a temperature in this region dominates the SED. The blackbody curve wth T=76,000K represents the temperature that would naively apply given the observed luminosity and mass of 3C~273 (see text). The dotted line labelled ``K97'' represents the ionising continuum used by \citet{Korista1997}. The normalisations of all components are arbitrary; they are scaled for comparison purposes.}
\label{fig:SED-comp}
\end{figure*}

Fig. \ref{fig:SED-comp} shows the SED of the nearby quasar 3C~273, plotted in linear units, compared to a blackbody of temperature T$\sim$32,000K which is a good first approximation to the observed SED.  According to \citet{Peterson2004} the mass  of the black hole in 3C~273 is 8.87$\times$10$^8M_\odot$. (Note that \citet{Kaspi2000} and \citet{Paltani2005} find values roughly a factor of two either side of this value.) Integrating over the SED shown in Fig. \ref{fig:SED-comp}, and extrapolating the FUV with $\alpha=-1.8$, the total luminosity is approximately L$_{BB}\sim$4$\times$10$^{39}$W, so that L$_E\sim$0.356.  If this luminosity were radiated from 5R$_S$, the expected black body temperature would be T$\sim$76,000K. This is clearly discrepant with the observed SED, as also shown in Fig.\ref{fig:SED-comp}.

One could fix the temperature problem by making AGN radiate well below the Eddington limit, but only by making their black holes almost two orders of magnitude more massive. Alternatively one could require that the radiation emerges from a larger radius - of the order 50$R_S$. Of course most of the energy generation has to take place interior to this radius, or once again an unreasonably large mass is required. One is led therefore to the possibility that interior energy is absorbed and reprocessed, and that the peak we see corresponds to the reprocessed rather than the original emission. 

The FUV SED shortward of the 1100\AA\ peak is more controversial. \citet{Zheng1997} and \citet{Telfer2002}, by compiling a composite SED using HST data spanning a large range of redshifts, find that the FUV falls steeply all the way to 300\AA\, as $S_\nu \propto \nu^{\alpha}$ with $\alpha\sim -1.8$. Constructing such a high-z composite requires careful correction for Ly$\alpha$ forest absorption by the intervening IGM. On the other hand, \citet{Scott2004}  use FUSE data for relatively low redshift AGN to construct a composite reaching 650\AA\ and find a mean slope of $\alpha\sim$-0.6. They find considerable FUV diversity, and a tendency for high-luminosity AGN to be steep, and low-luminosity AGN to be flat. \citet{Binette2005}, \citet{Shang2005}, and \citet{HaroCorzo2007} examine the FUV SEDs of individual AGN, up to z$\sim$1, where the IGM correction is still small, and confirm the diversity of FUV shapes. In particular, \citet{Binette2005} stress that while some quasars continue to decline in $\nu L_\nu$ out to 600\AA\, others turn back up at short wavelengths. Fig. \ref{fig:SED-comp} shows examples of these two cases taken from their paper.

Another way of altering the expected emission is by absorption. Various possibilities are reviewed by \citet{Binette2008}. The observed knee is close to the Lyman edge, but  the shape is not like an edge, and it is definitely at somewhat longer wavelength. It is also possible that dust reddening has altered the shape - de-reddening the 3C273 SED by  $A_V\sim 0.35$ over and above the line of sight Galactic extinction of $A_V\sim0.1$ makes the match with the expected temperature much better. \citet{Binette2005} explore the possibility that crystalline dust can produce a broad absorption trough matching that seen in PG1008+1319 and similar objects. Overall, absorption models remain plausible, but I do not consider them further in this paper.  


\subsection{The ionisation problem} 

The ionisation problem is a corollary of the temperature problem : an SED falling in $\nu L_\nu$ terms does not seem capable of producing the broad emission lines observed in AGN. This is especially true if the high-redshift composites derived by \citet{Zheng1997} and \citet{Telfer2002} are correct; these show a very soft FUV continuum, with $\alpha\sim -1.8$, whereas the observed emission lines in the optical and near-UV seem to indicate a much harder continuum,  with $\alpha\sim -0.5$.  Fig. \ref{fig:SED-comp} contrasts the observed SED with the ionising continuum used in the comprehensive BLR study of \citet{Korista1997} -  a power law with index $\alpha=-0.5$ and an exponential cut off at $kT$=44eV - which does a good job of reproducing the observed ratios of broad emission lines in AGN. The ionisation problem has been known for many years \citep{Netzer1985, Collin-Souffrin1986, Dumont1998, Korista1997a}. Netzer's (1985) argument concerned the ratio of the total BLR emission to Ly$\alpha$. Roughly speaking, if Ly$\alpha$ tells us the continuum near the Lyman edge, then extrapolating with a steep slope does not provide enough energy to produce all the BLR emission. Netzer argued that the problem could be solved if the BLR (but not the continuum) is reddened by $A_V \sim$ 0.25--0.5, but \citet{Collin-Souffrin1986} shows that the observed continuum then fails to produce the reddening-corrected line ratios. \citet{Korista1997a} concentrate on the HeII lines, especially HeII1640, and show that a steeply extrapolated continuum in the manner of \citet{Zheng1997}  fails to produce the observed HeII 1640 equivalent width by at least a factor of 5, and probably considerably more. Korista et al consider several possible solutions. One is that the UV-EUV bump is double peaked. This is the possibility that I pursue in this paper.

Fig. \ref{fig:SED-comp} illustrates the ionisation problem, contrasting the observed SED of 3C~273, the \citet{Telfer2002} composite, and the \citet{Korista1997} ionising continuum. Interestingly, although the observed SED seems inconsistent with both the required ionising continuum, and the naively expected temperature, the ionising continuum and the naively expected temperature are roughly consistent with each other. Furthermore the FUV upturn shown by PG1008+1319 is roughly consistent with the slope expected from the naively expected temperature, suggesting that this very hot emission is present, but somehow masked in the NUV region.

\subsection{The timescale problem} 


Relatively few AGN have well sampled UV light curves, but those that do show fairly well defined ups and downs.  The situation is summarised by \citet{Collier2001a}, who derive UV structure functions for four AGN (NGC 3783, NGC 7469, NGC 5548, and F9) and optical structure functions for six other AGN. They show characteristic timescales (defined by the flattening of the structure function) of between 5 and 94 days. Collier and Peterson show evidence that this characteristic  timescale is proportional to black hole mass, which for these UV-studied local AGN is of the order $10^7 - 10^8$M$_\odot$. This is roughly confirmed by the fact that much more luminous quasars seem to show timescales of the order of years \citep{Giveon1999, Trler1999, Trevese2001, Trevese2002, Hawkins2007}. 


A timescale of the order of tens of days is inconsistent with simple quasi-steady accretion disc models, as the viscous (radial drift) timescale is far longer, as was first recognised by \citet{Alloin1985}. Following \citet{Frank2002}, Ch 8, and scaling to the time expected for a radius appropriate to a standard thin disc radiating at a characteristic NUV temperature of T=30,000K, the viscous timescale is

\[ t_{visc} = 12.6 {\rm \, yrs} \times L_E^{-3/10} M_8^{6/5} R_{30}^{5/4} \alpha_{0.1}^{-4/5} \mu_{0.1}^{3/10} \] 

\noindent where $\alpha$ and $\mu$ are the usual viscosity and efficiency parameters, scaled to their expected values.  Acoustic modes also do not work. For T=30,000K, assuming the gas is largely ionised, the sound speed is $\sim$ 22 km s$^{-1}$, so that the sound crossing timescale is $t_{sound} = 12.6{\rm \, yrs} \times R_{30}  M_8$. On the other hand, the light crossing timescale,  $t_{light} \sim 8.3{\rm \, hrs} \times  R_{30} M_8 $ is much too short, although it may be relevant to the co-ordination of variability at different wavelengths, as discussed below.

There are two timescales that may work. The first is some kind of dynamical timescale. The free fall timescale from 30R$_S$ is  $t_{ff} \sim 5.9{\rm \, days} \times R_{30}^{3/2} M_8$. This is slightly on the fast side, but very much in the right ball park, suggesting that accretion flow may be dynamically unstable. The second possibility is some kind of cooling timescale, which will for example determine the time in which internally stored energy could be released as radiation in a flare. This is much more model dependent, depending on the total mass involved, its temperature, and whether it is optically thick or thin. \citet{Collier2001a} state that for a standard accretion disc model the thermal or cooling timescale is $t_{thermal} \sim 27.5{\rm \, days} \times  R_{30}^{3/2} M_8 \alpha_{0.1}^{-1}$ where $\alpha$ is the standard disc viscosity parameter.

If either a dynamical timescale or a thermal timescale is the correct answer, the variability cannot be modelled by a quasi-steady accretion disc - the physical structure of the disc is changing faster than the gravitational energy is being dissipated. However this problem could be removed if what we see is reprocessed emission - only the reprocessor has to change fast, not the disc itself. In section \ref{sec:timescales} we will look at whether dense reprocessing clouds can provide the right timescale.

\subsection{The co-ordination problem} \label{coordination}

The next serious problem for accretion disc models is that variations at different UV-optical wavelengths occur in phase - i.e. the peaks and troughs line up, so that any delay or smearing is small compared to the variability timescale itself, which is very hard to achieve if the radiation at different wavelengths come from different parts of the disc. The best evidence comes from a handful of local Seyfert galaxies with well sampled UV light curves (NGC 5548 : \citealt{Clavel1991, Korista1995}; NGC 4151 : \citealt{Edelson1996, Crenshaw1996}; NGC 7469 : \citealt{Kriss2000}; NGC 3516 : \citealt{Edelson2000}). For more luminous quasars, there are few UV datasets, but fairly well sampled long term optical light curves presented by Giveon \et (1999) and Hawkins (2003) show changes at different wavelengths that are also aligned to much less than the typical variability timescale.  The lack of wavelength-dependent delays led to the suggestion that the inner accretion disc is heated at least in part by the central X-ray source (rather than locally by viscous torques), so that the UV variations are actually a reprocessed version of variations in the central X-ray source  \citep{Krolik1991}. The expected delay between different wavelengths is then the light travel time between different regions, which is of the order hours. Comparing UV wavelengths, 
several authors have claimed a measurement of such a short delay for NGC 7469 \citep{Wanders1997, Collier1998, Kriss2000}, but for other AGN (NGC 5548, 3516, 3783, 4151, and Fairall 9), attempts have only resulted in limits of the order 0.1 to 0.3 days  \citep{Crenshaw1996, Edelson1996, Edelson2000, Peterson1998}. \citet{Korista2001} have claimed that the NGC 7469 effect is an artefact due to contamination of the continuum by diffuse emission from the BLR. Comparing different optical wavelengths, a number of authors have shown marginally significant delays of the order of days in several different AGN \citep{Collier2001, Oknyanskij2003, Sergeev2005, Doroshenko2005, Breedt2009}.




A prediction of the ``disc heated by X-ray source'' picture is that the UV and optical emission should lag the X-ray emission. However, simultaneous monitoring campaigns show that any such delay is less than 0.15 days in both NGC 4151 \citep{Edelson1996} and NGC 3516 \citep{Edelson2000}, and in the case of NGC 7469, if anything the UV {\em leads} the X-ray emission by $\sim 4$ days. \citet{Breedt2009}, from 5 years of monitoring MKN 79, find that overall, X-ray and optical emission are well correlated, with an optical vs X-ray lag of less than $\sim$2 days, but also note differences. The X-ray variability has more short timescale power, as might be expected, but when smeared with the expected disc transfer function it fails to reproduce the observed optical light curve (see their Fig. 7) - the predicted light curve matches some but not all of the observed flaring on timescales of tens of days, and does not produce the long term trends on timescales of hundreds of days. Similar problems were noted by \cite{Kazanas2001}. 

A further problem is that the X-ray luminosity in AGN is much smaller than the the UV luminosity. For relatively low luminosity AGN, the total X-ray luminosity is sensitive to how the observed 2-10 keV spectrum is extrapolated, so that X-ray luminosity may in fact be comparable to the UV luminosity \citep{Chiang2003}. However, this cannot be a universal explanation, as the X-ray/UV ratio is an order of magnitude smaller for the most luminous quasars \citep{Strateva2005}. Overall there is clearly a close connection between UV-optical emission and X-ray emission, but the relation is a complicated one. One possibility is that, rather than the X-ray emission being primary and the UV a reprocessed version, the real primary emission is the FUV peak at $\sim$ 300\AA\  which is predicted by simple accretion disc theory and photo-ionisation calculation.  We do not have FUV light curves, but one AGN, NGC 5548, has been monitored the other side of the possible peak, at $\sim$ 80\AA\ with EUVE. \citet{Marshall1997} showed that over ten days the variations at 76\AA\ and 1350\AA\ are well correlated, with any lag less than 0.25 days, but that the amplitude of variations was several times larger in the EUV. In a later EUVE observation made simultaneously with RXTE and ASCA, \citet{Chiang2000} showed that the 76\AA\ emission leads the X-rays by 0.4 days.  There is a case to be made then that {\em both} the UV and X-ray emission are substantially reprocessed.





\subsection{The size problem} \label{size}

Sizes deduced from the fluctuations of gravitationally lensed quasars seem to be discrepant with expectation. The light from each lensed component can suffer additional micro-lensing by stars, or possibly by dark matter substructure. The expected optical depth is close to unity, so that each component has ongoing irregular fluctuations that are uncorrelated with each other, and can therefore be distinguished from intrinsic variability common to all components. (See for example the review of \citealt{Wambsganss2006}.)  The magnification depends on source size, as the expected size, while small compared to the Einstein radius of the lenses, is not point like - i.e. a larger source has more washed out fluctuations. A clear qualitative result is that component-to-component anomalies are significantly bigger in X-rays than in the optical-UV, showing that the X-ray source is more compact, by  a factor of several \citep{Dai2010, Chartas2010} and possibly that the hard X-ray size is smaller than the soft X-ray size \citep{Chen2012}. Likewise, broad emission lines show little or no fluctuation, and therefore the BLR size is substantially bigger than the continuum source size \citep{Lewis1998}. Within the optical-UV range, fluctuations are larger at shorter wavelengths, (e.g. \citealt{Eigenbrod2008, Mosquera2011}),  strongly supporting the general idea of a distributed source with a colour gradient. The slope is plausibly consistent with the radial temperature gradient expected from an accretion disk, but with a rather large error. This is an area that will hopefully improve strongly in coming years.

As well as these striking relative results, lensed quasar monitoring can give us information on the absolute source size. This is somewhat model dependent as it requires understanding the macroscopic lensing, the mass function of the microlenses, and the relative transverse motions. Early results claimed either consistency with expected source sizes \citep{Wambsganss1990} or that observed sources sizes were smaller than expected accretion discs \citep{Rauch1991}. More recent and substantial studies have consistently concluded that observed sizes are a factor of several larger than expected from simple thin accretion disk models of the same objects \citep{Pooley2007, Morgan2010, Jimenez-Vicente2012}, while correlating clearly with the black hole mass estimated from line widths \citep{Morgan2010}. Suggested solutions to this problem could include much lower accretion efficiency, or a flatter temperature profile \citep{Morgan2010} or an inhomogeneous accretion disc \citep{Dexter2011}. The central reprocessing region discussed later in this paper is another possible solution to this problem.

\subsection{Amplitude effect and mixing model} 

\begin{figure}
\centering
\includegraphics[width=0.35\textwidth,angle=-90]{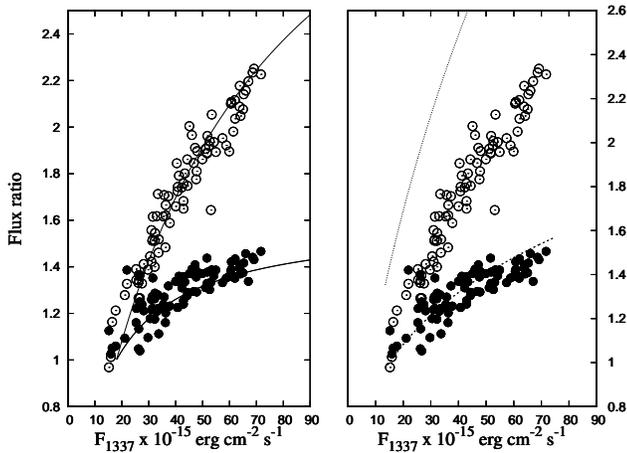}

\caption{\it\small Mixing models for the colour-amplitude trend in the variability of NGC 5548. The data are from \citet{Clavel1991}. The open circles show the flux ratio F1337/F2670, and the filled circles show F1337/F1825, where F1337 refers to F$_\lambda$ at 1337\AA . The model tracks show the effect of a fixed power law and a variable black body. In the left hand pane, the temperature of the black body is fixed at T=27,500K, with only its normalisation varying. In the right hand pane, the emitting area of the black body is fixed and its temperature varied. }
\label{fig:mixing}
\end{figure}

Related to the co-ordination problem is the fact that the amplitude of variability for both local AGN and luminous quasars
is strongly wavelength dependent \citep{Cutri1985, Clavel1991, Wilhite2005}. This is most clearly demonstrated in Fig. 13 of \citep{Wilhite2005}, which shows the ratio of the mean difference spectrum to the mean spectrum for 2500 SDSS quasars. This variability index increases steeply at wavelengths shorter than 2500\AA . A generic way to explain the amplitude and co-ordination effects at the same time is by mixing a constant red component with a variable blue component \citep{Lawrence2005}. However, only some such models work. Fig. \ref{fig:mixing} shows two deliberately simple toy models applied to the NGC 5548 campaign data of \citet{Clavel1991}. The constant component is modelled as a power law with slope $\alpha$ and normalisation $N_{PL}$,  and the variable component as a blackbody at a single temperature $T$ with normalisation $N_{BB}$. 

In the first model, the temperature $T$ of the black body was fixed, and flux-ratio values were calculated for a sequence of values of the black-body normalisation $N_{BB}$, to produce predicted tracks in two diagrams - the ratios F1337/F1825 and F1337/F2670 {\em vs} the flux F1337, where F1337 refers to the flux at 1337\AA\ .  The other parameters  ($N_{PL}$, $\alpha$, and $T$) were adjusted to produce a good simultaneous (eyeball) fit between these tracks and the data. The best fit temperature for the variable black body was T=27,500K, with the power law of the fixed component having slope $\alpha$ = -2.0.

In the second model, the emitting area of the black body was held fixed, and the temperature $T$ varied to produce the predicted tracks. This of course produces variation in both colour and normalisation.  As with the first model, the free parameters  ($N_{PL}$, $\alpha$, and black body emitting area) were adjusted to try to achieve a simultaneous fit to the two flux ratios, but no fit could be found. As an illustration, Fig. \ref{fig:mixing} shows parameters adjusted to fit the run of the ratio F1337/F1825, which results in a prediction for F1337/F2670 which is a long way off.

Of course both models are physically unrealistic - the variable component is unlikely to be a single blackbody, the red component may not be well represented by a power law, and the variable component may vary in both area and temperature. Nonetheless, the difference between the two simple extremes is striking. It strongly suggests that the way to explain the variability of NGC~5548 is by the variability of a component with {\em fixed shape}. 


\section{Reprocessing by dense inner clouds}  \label{sec:clouds}

\subsection{Introduction}

The temperature problem (i.e. the premature turnover of the SED) and the variability behaviour of AGN suggest that the observed UV continuum represents re-processed emission. The earlier idea that the UV represents reprocessed X-rays seems untenable because there is not enough X-ray luminosity, and because in at least some cases the X-ray emission lags the UV emission. The primary radiation could instead be the unseen peak EUV emission - the inner disc heating the outer disc. Against this idea is the observation of steeply declining FUV continuum. In support of the idea is the fact that emission lines suggest that there is indeed a hard EUV continuum, and the fact that at least some high redshift quasars show a turn up shortward of 800\AA . Perhaps, as suggested by \citet{Korista1997}, the continuum is double peaked.

However, there remains the puzzle that the wavelength of the $\nu L_\nu$ peak ($\sim$ 1100\AA ) seems to be universal \citep{Shang2005}, even if the continuum a few hundred \AA\ shortward of this is not \citep{Scott2004, Binette2005, HaroCorzo2007}. Furthermore, within individual objects, it seems that multi-wavelength changes may be explained by the variability of a component of fixed shape. Where a spectral shape is relatively constant across a large range of conditions, this is suggestive of a solution dependent on atomic physics rather than macroscopic physics. Perhaps what we see as the UV Bump is actually made of emission lines. The features that we recognise as the ``broad emission lines'' come characteristically from a radius of $\sim$1000R$_S$. Clouds at say $\sim$30R$_S$ would produce emission lines 6 times broader, which would be seen as a false continuum. A similar logic has led some previous workers to consider optically thin (primarily free-free) emission from hot (10$^{5-6}$)K gas as an alternative to the accretion disc \citep{Ferland1990, Barvainis1993}, but my aim here is to discuss the role of optically thick clouds in conjunction with the accretion disc.  In the next subsection, I discuss some general issues concerning dense inner clouds. Following this, I show model results for emission from such clouds, how they could produce a false local continuum peak at 1100\AA\ , and discuss the prospect for more detailed models. In section \ref{sec:discuss} I discuss whether we expect such clouds to exist, and how this relates to other issues such as the X-ray spectra of AGN.

\subsection{Cold dense clouds}\label{sec:oldclouds}

Cold dense clouds in AGN were a minor fashion for just over a decade \citep{Guilbert1988, Ferland1988, Lightman1988, Celotti1992, Sivron1993, Collin-Souffrin1996, Kuncic1996, Kuncic1997, Czerny1998, Celotti1999}.  In all these papers, it was assumed that the primary energy source was either non-thermal (with power law slope -1) or a very hot X-ray plasma, with the issue being whether ``cold'' thermal material could survive within the non-thermal plasma, and whether the reprocessed thermal emission could be a viable alternative to an accretion disc for explaining the optical-UV emission in AGN. These ideas seem to have gone out of fashion because it has become widely accepted that an accretion disc is the primary mode of energy generation. An overlapping series of papers has considered the effect of such cold clouds on the X-ray spectrum, especially the Fe line and the Compton reflection hump \citep{Lightman1988, Sivron1993, Karas2000, Malzac2002, Merloni2006}, and also the possibility that variable partial covering explains X-ray variability \citep{Abrassart2000}. However this has been a minority industry, as most workers have followed \citet{Pounds1990} in assuming that X-ray reflection is from the surface of the accretion disc.

My approach in this paper differs from the above papers in two respects. Firstly, rather than trying to produce an alternative to the accretion disc, I will examine how cool dense clouds can modify the accretion disc SED. Secondly, I will concentrate on a somewhat different region of parameter space. Most of the above papers consider very dense clouds in the very central regions (3R$_S$$<$R$<$10R$_S$) where most of the energy is generated, arguing that they can be confined by the strong magnetic fields that may be present. \citet{Kuncic1996} argued that such clouds must be very small and dense (hydrogen density n$\geq$10$^{14}$ cm$^{-3}$R$_3^{-2}$M$_8^{-1}$) if they are to cool in less than the free-fall time. The radiation from such clouds will be close to black body at around 10$^5$K, as shown by detailed calculations in nearly all the above papers, and as such would be hard to distinguish from other forms of cold material, such as an accretion disc, and indeed this emission would suffer from many of the same problems previously discussed in section \ref{sec:probs}. However, at somewhat larger radial distances, say 30R$_S$, the density required to cool in a reasonable time is much less, n$\sim$10$^{12}$. Such clouds will radiate mostly in atomic transitions, but they will still be moving fast enough that the Doppler blurred emission will produce a false continuum. If the accretion disc itself is the source of such material, then there is no confinement issue, and the starting temperature of the clouds ($\sim$10$^4$K) will be similar to the expected photo-ionisation equilibrium temperature when exposed to the $\sim$10$^5$K radiation from the inner region.  I will consider the origin of the clouds a little more closely in section \ref{sec:discuss}. For now, I will assume that moderately dense clouds exist at R$\sim$30R$_S$, and consider the consequences. I note that \citet{Czerny1998} have also examined n$\sim$10$^{12}$ clouds,  but they consider clouds at rather higher ionisation parameter, and assume a non-thermal primary continuum.

\subsection{Emission from cold dense clouds}

I examined the properties of clouds ionised by a hot central source at a 3C~273-like luminosity, using version C08.00 of the well known ``Cloudy'' software package \citep{Ferland1998}. Except where noted below the ionising source was a blackbody with temperature $T$=100,000K and (total) luminosity L=4$\times$10$^{46}$erg s$^{-1}$, and the distance of the clouds from the source was $R$=10$^{16}$cm, corresponding to 30R$_S$ for M$_H$=10$^9M_\odot$. The calculations were performed assuming an open geometry. A variety of densities and column densities were examined, as described in the subsections below. The temperature of 100,000K could be seen as an approximation to the disc emission interior to the clouds, or if perhaps the disc is disrupted, could be thermal emission from the very dense interior clouds discussed in section \ref{sec:oldclouds}. In many of the cases described below, I tried varying the temperature over the range 75,000 to 200,000, at fixed total luminosity, and found no qualitative difference. In a few cases, I also tried a continuum which added a two component X-ray continuum to the 100,000K black body, based on the data in Kriss et al (1999). I used a soft X-ray power law with slope $\alpha$=-1.5, a low energy break at 7.5 Rydberg,  and luminosity L=5$\times$10$^{45}$erg s$^{-1}$; and a hard X-ray power law with $\alpha$=-0.6 and luminosity L=1.25$\times$10$^{46}$erg s$^{-1}$. The main effect of such an X-ray continuum was to turn a back-end neutral zone into an extended partially ionised zone. The effect on the reflected spectrum described below was significant but not large - around 20\%.  Finally, the choice of 30R$_S$ is guided by the logic described above in section \ref{sec:oldclouds}, but I also considered some clouds at other distances, as described below. 

In the subsections below I refer to $\log{n}$ and $\log(N_H)$ where the hydrogen number density $n$ is in units of cm$^{-3}$ and the hydrogen column density N$_H$ is in units of cm$^{-2}$.

\subsubsection{BLR-like clouds, $\log{n}\sim 10$}

Clouds with density $\log{n}$=10, such as are often assumed in BLR models, can be considered for column densities $\log{N_H}<$24.5. Above this $N_H$ value, the cloud size is larger than the whole region. Throughout this column density range, the clouds are fully ionised in H and He. At low columns, $\log{N_H}<23$, the clouds are essentially transparent. Above this range, the electron scattering depth becomes significant, and an increasingly large fraction of the continuum is reflected back. The total emission, which is what one would see if  equal numbers of front faces and rear faces are visible, is not very different from the input continuum. There are significant emission lines, but not strong enough to produce the effect we are looking for here. Clouds at this density may of course exist in the inner region, and produce observable absorption and emission lines, but will not produce a gross modification of the SED.


\begin{figure*}
\centering
\includegraphics[width=0.5\textwidth,angle=-90]{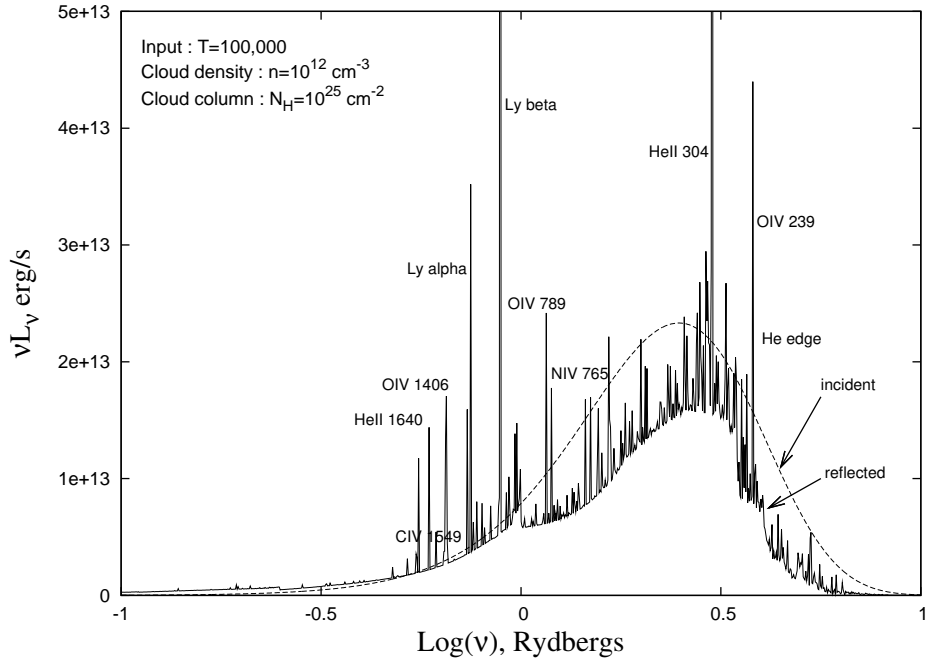}
\caption{\it\small The emitted spectrum from dense inner clouds at a distance 10$^{16}$cm from a blackbody radiation source of temperature 100,000K with luminosity 4$\times$10$^{46}$ erg s$^{-1}$. The vertical units are erg s$^{-1}$ cm$^{-2}$. Only the incident and reflected spectra are shown, but the transmitted and diffuse-out components are very small. The line-to-continuum contrast is approximately what would be seen at velocity resolution 1000 km s$^{-1}$. }
\label{fig:cloudspec}
\end{figure*}


\subsubsection{Dense clouds, $\log{n}\sim 12$}\label{n12-clouds}

Clouds with density $\log{n}$=12 are much more promising. At low column densities, they are ionised, with the results being fairly similar to lower density clouds and with little net effect on the SED. However  at $\log{N_H}$=24.55 and above the rear part of the cloud becomes neutral. (With X-ray continuum added, this rear zone is partially ionised, but the transition is in the same place.) 
As the total column increases above this value, the ionised column stays the same and the neutral column simply increases. The ionised column corresponds to the point where the electron scattering optical depth is 2.3. Varying the density from $\log{n}$=11 to 13 changes the transition optical depth a little, from 4.0 to 0.9. 

The emission from the cloud is more or less the same for all columns $\log{N_H}>$24.6, and is very asymmetric.  Seen from the far side, the cloud is essentially opaque at all wavelengths. However, the cloud is a strong reflection source in both continuum and lines. The resulting spectrum is shown in Fig. \ref{fig:cloudspec}. The electron scattering depth of the ionised zone is of order unity, so that a large fraction of the continuum is reflected back. However around 8\% of the incident EUV continuum is reprocessed into emission lines,  the strongest lines being He II Ly$\alpha$ at 304\AA\ and HI Ly$\beta$ at 1025\AA . The summed luminosity of the lines in the 1000\AA\ region is $\sim$30\% of the incident luminosity at $<$1Ryd. This, together with the marked drop shortward of the Lyman edge, is sufficient when velocity-blurred to make a substantial but localised false continuum, as discussed in section \ref{sec:smearmodels}.

\begin{figure}
\centering
\includegraphics[width=0.33\textwidth,angle=-90]{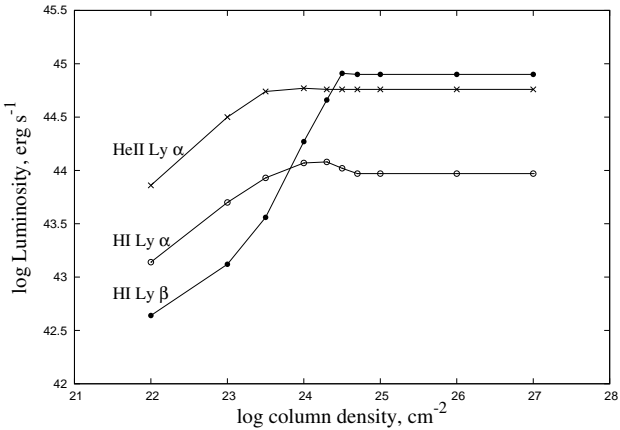}

\caption{\it\small Dependence of reflected line luminosity on column density. In all cases the other parameters are $\log{n}$=12, $T$=100,000K, $\log{L_{total}}$=46.6 and $\log{R}$=16. The luminosities are calculated on the basis of a 4$\pi$ covering factor.}
\label{fig:col}
\end{figure}

Over the range 22$<\log{N_H}<$24.5 the luminosity in Ly$\beta$ increases by more than two orders of magnitude, as shown in Fig.\ref{fig:col},  while the ratio Ly$\beta$/Ly$\alpha$ increases from $\sim 0.1$ to $\sim 8$, and the ratio Ly$\beta$/He II Ly$\alpha$ increases from $\sim 0.1$ to $\sim 1.4$. From $\log{N_H}$=24.5 to the maximum plausible column of  $\log{N_H}$=27, the luminosity in Ly$\beta$ is constant, and the ratios Ly$\beta$/Ly$\alpha$ and Ly$\beta$/He II Ly$\alpha$ are constant at 8.51 and 1.38 respectively. If we fix the column density at $\log{N_H}$=25 and vary the temperature of the ionising source from T=75,000 to T=200,000, the ratio Ly$\beta$/Ly$\alpha$ changes little (from 8.71 to 8.32), but the ratio  Ly$\beta$/He II Ly$\alpha$ changes from 4.27 to 0.45. This last sensitivity may give a way to test models of this kind. 

\begin{figure}
\centering
\includegraphics[width=0.3\textwidth,angle=-90]{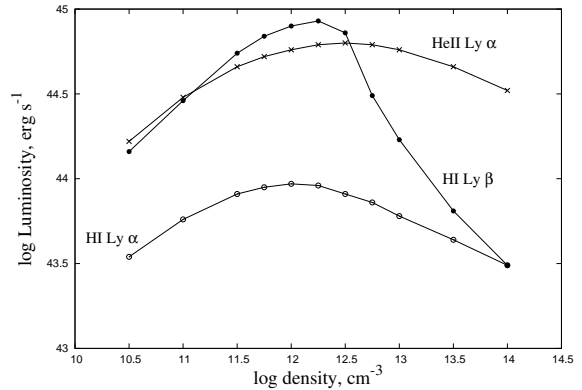}

\caption{\it\small Dependence of reflected line luminosity on density. In all cases the other parameters are $\log{N_H}$=25, $T$=100,000K, $\log{L_{total}}$=46.6 and $\log{R}$=16. The luminosities are calculated on the basis of a 4$\pi$ covering factor.}
\label{fig:density}
\end{figure}

Clouds with $\log{n}$=11 give a similar result. Fig. \ref{fig:density} shows how the luminosity in various lines changes as the density is varied, for  $\log{N_H}$=25 and T=100,00K.  The Ly$\beta$ effect is significant over a range of density, but strongly peaked around $\log{n}\sim$12.

Similar clouds much closer in give a somewhat different result. At 5$R_S$, a cloud with $\log{n}$=12 and $\log{N_H}$=25 still has a neutral back end, and so is opaque when seen from the rear, but the larger ionised column results in larger electron scattering depth, $\tau_{es}=5.3$, so that the reflected continuum is little different from the incident continuum, and the reflected lines are a order of magnitude smaller than the 30$R_S$ case. Further out, similar clouds at 100$R_S$ have $\tau_{es}$=0.7 and give a result similar to the 30$R_S$ case, with Ly$\beta$ in fact several times stronger, although of course these lines will be distincly less broad than those at 30$R_S$. By 300$R_S$, line reflection is still strong, but now dominated by Ly$\alpha$ rather than Ly$\beta$.


\subsubsection{Very dense clouds, $\log{n}\sim 14$}

What about clouds with even higher density, say $\log{n}>14$? This strays into the region which Cloudy was not designed for, and so issues warnings, but the rough results are clear. At low columns, such clouds are ionised, but at $\log{N_H}>$23
they are mostly neutral with a thin ionised skin. There is little continuum reflection; the reflection spectrum is dominated by HI Lyman continuum emission, with He I Lyman continuum and He II Ly$\alpha$ also strong, and significant HI Balmer continuum. The relative strength of these components seems to be roughly constant at $\log{N_H}>$24. I do not explore these densities in more detail, because the results from Cloudy may be unreliable and also because they do not seem to give the 1100\AA\ peaking that observations require. 

At even higher densities, clouds will of course start to look like black bodies, and emit at a temperatures in the range 10$^{4-5}$K, depending on radial distance, as discussed by several previous authors (see discussion in section \ref{sec:oldclouds})

\subsection{Prospect of dense cloud reflection models}\label{sec:smearmodels}

Adding velocity blurring to the reflected spectrum of a single cloud can produce a false continuum of roughly the right kind. What about an ensemble of clouds surrounding the accretion disc - will the emission be dominated by the intrinsic continuum or the reflected spectrum ? This will depend on both the covering factor of the clouds, and the amount of blurring. I blurred the spectrum shown in Fig. \ref{fig:cloudspec} with a Gaussian of FWHM 75,000 km s$^{-1}$, roughly what one might expect for clouds at 30$R_S$, depending on the kinematic model. The reflected quasi-continuum $\nu S_\nu$  at 1100\AA\ is then roughly twice the incident continuum at that wavelength. If the covering factor is $C$ then the amount of leaked-through intrinsic continuum is proportional to $1-C$ and the fraction of the total due to reflection, ignoring self-occultation effects, is then approximately $2C/(1+C)$. For the reflection spectrum to dominate therefore requires $C>1/3$. The fraction changes quite quickly around this value, with C=0.1 giving 18\%, and C=0.5 giving 67\%. Larger covering factors require self-occultation to be taken into account. There is therefore only a small range of covering factors which produces a strong reflection spectrum without self-occultation.

\begin{figure*}
\centering
\includegraphics[width=0.5\textwidth,angle=-90]{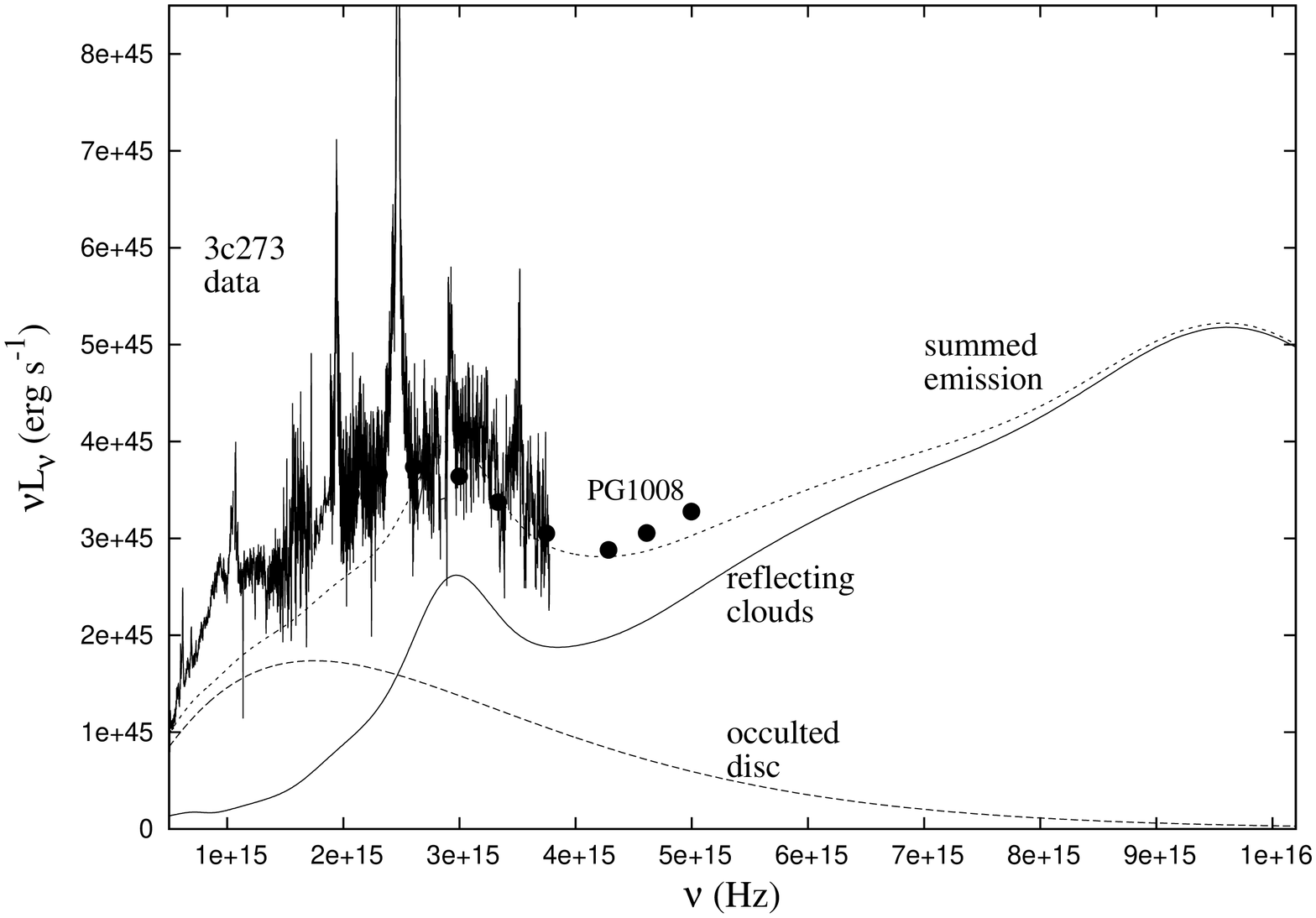}

\caption{\it\small An illustrative model, compared to the combined SED of 3C~273 \citep{Kriss1999} and PG 1008+1319 \citep{Binette2005}. The datapoints for PG1008+1319 are shifted vertically to match onto the 3C~273 data. The black hole is assumed to have mass $M_9=0.89$ and to be radiating with $L_E=0.36$. The  reflecting clouds are assumed to have  $\log{n}=12$ and $\log{N_H}$=25, to be located at 30$R_S$ and to be ionised by an intrinsic continuum with temperature $T=100,000$K. The reflection spectrum is smeared with a Gaussian of FWHM 75,000 km s$^{-1}$. The power law follows $\nu^{1/3}$ with an exponential cut-off at 0.4 Rydberg, roughly corresponding, for our parameters, to the expected accretion disc occulted at 30$R_S$.
}
\label{fig:smearedspec}
\end{figure*}

However self-occultation, with large covering factor, may be exactly what we need. Note that the reflected spectrum from a single cloud, like the incident continuum, is dominated by emission at several Rydbergs. Clouds which see only other clouds, rather than the disc surface, will also produce a spectrum like that of Fig. \ref{fig:cloudspec}. The cloud spectrum bounces from one cloud to another and emerges at the edge of the system of clouds at a kind of quasi-photosphere. The central disc will be completely occulted, so the SED that we see is composed of the pure cloud spectrum plus the un-occulted outer part of the disc. Fig. \ref{fig:smearedspec} shows a toy model of this kind compared to the combined data for 3C~273 and  PG 1008+1319.  The model shows a blurred version of the Fig. \ref{fig:cloudspec} reflection spectrum, added to  a $\nu^{1/3}$ power law with an exponential cut-off at 0.4 Rydbergs, meant as an approximation to an accretion disc truncated at $30 R_S$. The relative strengths of the two components are roughly correct, but note that this is not a full or self-consistent model. It should be seen as  a plausibility demonstration rather than a real model. Significant extra work, and decisions on model assumptions, is needed before it makes sense to attempt a proper fit. 

(i) There will very likely be a wide range of densities and column densities at different radii. To some extent, the SED modification may naturally be dominated by clouds in a particular range - by comparison with these just--right clouds, clouds of lower density or column may be transparent, and clouds that are closer in may primarily reflect an unaltered continum.  However, a prediction of the actual range of cloud parameters could make a big difference.

(ii) The shape of the SED modification will be sensitive to the radial distribution of velocity and cloud density. In this paper, I make basic arguments about the likely density and velocity regime concerned. Going beyond this needs a physically motivated model which makes a prediction of the velocity and density distributions.

(iii) A realistic model for the intrinsic continuum is also needed, both for the interior region, which provides the continuum which ionises the clouds, and the exterior region, which produces the optical SED. The behaviour of the inner accretion disc is clearly crucial.

(iv) Geometrical considerations could also make a big difference. For example, it may be that clouds lifted out of the disc in a narrow range of radii will form a kind of funnel, somewhat akin to the model of Elvis (2000) but much closer in. In this case, the continuum that we see will be inclination dependent.

(v) To fit the whole optical-UV-FUV SED, we may need to take into account a more continous system of clouds, between my suggested $\sim$30R$_S$ region and the traditional $\sim$1000R$_S$ BLR region.


\section{Discussion}  \label{sec:discuss}

Do we expect that such dense thick clouds will exist in the inner regions of quasars ? Here I briefly discuss the mass of the possible cloud system, its origin, the lifetime of the clouds, the origin of UV variability, and the implications for AGN spectra. 

\subsection{Size of clouds and overall cloud system}

 Clouds with density n$_H$=10$^{12}$ cm$^{-3}$ and column N$_H$=10$^{25}$ cm$^{-3}$ will have thickness D$\sim$10$^{13}$cm. If spherical, they will have individual masses of the order 10$^{-6}$M$_\odot$, but of course it is quite possible that the material is the form of much more extended sheets or filaments. The total mass of the system of clouds can be estimated by considering a spherical sheet at radial distance $R$ with covering fraction $C$, which gives 

\[ m_s = 8.3 M_\odot \times C M_9^2 R_{30}^2  N_{25} \]

\noindent where as usual $M_9$ is black hole mass in units of 10$^9$M$_\odot$, R$_{30}$ is the radial distance in units of 30R$_S$, and N$_{25}$ is in units of 10$^{25}$ atoms cm$^{-2}$.  How does this compare with the mass likely to be involved in the accretion flow onto the black hole ? Let us first consider simple uniform spherical accretion. Assuming steady flow in free-fall the density at $R$ is 

\[   n = 2.2\times 10^8 {\rm cm}^{-3} \times \mu_{0.1}^{-1} M_9^{-1} L_E R_{30}^2      \]

\noindent where $L_E=L/L_{Edd}$, and $\mu$ is the usual radiative efficiency scaled to an expected value of 0.1. The accreting mass interior to $R$ is

\[   m_a = 1.1 M_\odot \times \mu_{0.1}^{-1} M_9^2 L_E R_{30}^{3/2}     \] 

\noindent The mass available is therefore an order of magnitude less than the required mass of clouds, and a very large degree of clumping would be needed to make  clouds of the necessary density, so it seems that a simple spherical accretion model does not work. 

There are two ways to increase the mean density and mass available. The first is for accretion to be inefficient ($\mu\ll 1$), as in ADAF models, where most of the energy generated by infall is advected into the black hole. Potentially a clumpy ADAF model could be constructed \citep{Celotti1999}, but I do not consider this further here as it seems unlikely to be relevant to luminous quasars. 

The second way to increase the mass available is to slow down the infall and so increase density by a factor $K=v_{ff}/v_{inf}$. In standard accretion disc models $K \sim \alpha^{-1} \left( R/H  \right)^2$ where $H$ is  the characteristic scale height of the disc and $\alpha$ is the usual viscosity parameter (\citealt{Frank2002}, Chapters 5 and 8). The density is further increased by the vertical compression factor $R/H$. For an AGN disc, assuming $\alpha \sim 0.1$ and $R/H \geq 10$, then the mass available within $R$ is at least $m_a\sim$ 10$^3 M_\odot M_9^{-1} L_E R_{30}^{3/2}$ and the mean density in the disc at $R$ is at least $n \sim$ 10$^{12} M_9^{-1} L_E$ cm$^{-3} R_{30}^{-3/2}$, with each of these quantities likely to be orders of magnitude larger. In conclusion there is a plentiful supply of material in the disc. We look at disc models a little more closely in the next section.

\subsection{Possible disc origin of clouds }

Detailed disc models (eg \citealt{Hubeny2000}) confirm the above rough calculation. From the figures in \citet{Hubeny2000}, I estimate that  for a disc radiating at 15\% of Eddington around a black hole of mass 10$^9$M$_\odot$, the mass interior to 30R$_S$ is  7$\times$10$^3$M$_\odot$, so only a very small fraction of the disc mass is needed. The density varies greatly, from $\sim 10^{15}$ cm$^{-3}$ in the midplane to  $\sim 10^{10}$ cm$^{-3}$ at the disc surface. Likewise the total column density is of the order $10^{28}$ cm$^{-2}$ or more, depending on radius, and thinning out towards the surface. Production of clouds then does not require complete disruption of the disc,  but just a clumping of the upper parts. It has been argued that in the inner regions discs become unstable to clumping (Krolik 1998) but in fact only a modest degree of clumping would be required to produce the necessary clouds.


To provide the reflection medium we are considering, the clouds also need to be lifted out of the disc, and to be cool enough to be not already completely ionised, i.e. a few times 10$^4$K. Discs are expected to have a significant vertical temperature gradient, so again this means that the upper portions would be the source of cloud material, rather than the disruption of the whole disc. In disc wind models (see eg \citealt{Proga2007} and references therein), uplift and gas temperature are closely linked. Radiation pressure comes from absorption of resonance lines as well as electron scattering, so that even well below the nominal Eddington limit gas can be lifted above the disc surface, where it is then hit by the central UV source and driven out. In the very inner disc, it is generally argued that the material is too highly ionised for this effect to work, so that winds are launched at $\geq 100R_S$, and only for high mass and high luminosity systems (eg \citealt{Shlosman1985, Murray1995, Proga2000}). However, \citet{Proga2005} shows that in the inner regions a ``failed wind'' or ``puffed up disc'' can be produced, which may be exactly the kind of structure we need.  Note that local radiation force is very sensitive to clumping. If clumps can form within the upper disc with the kind of parameters we have considered here, $\log{n}$=12 and $\log{N_H}$=25, they will be optically thick to the Lyman continuum, so that the ``force multiplier'' will be extremely large. Such clumps are then very likely to be ejected from the disc when formed. A reasonable guess is that at large radii ($>100 R_S$) gas can be uplifted without clumping, and so form a smooth wind, whereas at very small radii, ($<10 R_S$)  the material is too dense and hot to either feel uplift forces from heavy element absorption lines, or to make the cool thick clouds needed to get Lyman uplift. The result may be that dense clouds are formed and uplifted at a characteristic radius around 30R$_S$. More detailed calculations are needed to see if this is indeed the case.



Alternatively, clouds maybe uplifted by magneto-centrifugal effects, as in the models of \citet{Konigl1994},  \citet{DeKool1995}, and \citet{Everett2005}. This idea is usually considered at much larger radii, but may also be relevant in the inner regions.

\subsection{Cloud timescales and variability }\label{sec:timescales}

Do we expect such clouds to survive ? Unless confined (eg by an external medium or magnetic field) they would disrupt on the sound crossing timescale. Assuming that the gas is mostly ionised and at a temperature $T_{30}=T/30000$ this gives $t_{disrupt}=52.0 {\rm\ days}\times N_{25} n_{12}^{-1} T_{30}^{-1/2}$, which is very similar to the dynamical timescale at 30$R_S$. Because there is a plentiful supply of material, we do not require the clouds to be long lived. The cloud sound crossing timescale will also be (very roughly) the timescale on which clumps could grow inside the disc, so we may expect that clouds are repeatedly forming and being destroyed. 

How quickly would such clouds be expelled ?  If the clouds originate in the disc, they will initially be moving at orbital velocity, but because they are optically thick to Lyman continuum and electron scattering, they will be driven radially outwards with a greatly super-Eddington force. Given that they have mass per unit area $N_H m_H$, the time to reach escape velocity will be $T_{esc}=124 {\rm\ days} \times N_{25} M_9 R_{30}^{3/2} L_E^{-1}$. It seems therefore that clouds will be destroyed faster than they are expelled, and so we do not necessarily expect a wind to form - more of a broiling cloudy atmosphere. 

The covering factor of clouds may well vary significantly on the cloud formation/destruction timescale, leading to a natural explanation of UV variability. The intrinsic UV continuum, and the optical continuum, could vary quite slowly; it is just the reflected continuum that varies rapidly. The timescale is about right for the observed UV variability, and the fixed shape of the reflection spectrum is consistent with the strongly wavelength dependent variability, and the lack of delays between wavelengths. There could still be hours-timescale light travel time delays, just as in the disc reprocessing model. A possible objection to the idea that the intrinsic disc emission varies slowly is that the BLR emission lines, which track the observed UV continuum with a time lag, must surely be responding to the EUV continuum, which then must also be varying rapidly. However the dense inner clouds also reflect EUV continuum; the BLR clouds will see a mixture of slowly varying intrinsic EUV continuum and rapidly varying reflected EUV continuum. Whether this works quantitatively should be explored.

\subsection{Effect on X-ray spectrum and variability}

Much has been written about absorption and reflection effects in X-ray spectra (see the review by \citealt{Turner2009}). Ultimately a self consistent model for both UV and X-ray emission is needed, which is beyond the scope of this paper. I restrict myself to a few general points.

(i) In the picture emerging in this paper, the inner accretion disc is assumed to be present, so blurred ionised disc reflection will certainly occur. However, the clouds we have proposed will {\em also} produce blurred X-ray reflection effects, albeit at somewhat lower velocity blurring than the inner disc effects. If the inner disc is occulted, the cloud reflection effects may dominate. 
The clouds will also produce opaque partial covering, which can easily lead to an {\em apparent} reflection efficiency greater than 1. \citet{Merloni2006} have considered inhomogeneous X-ray reflection models, using dense clouds with no central accretion disc, but embedded in a very hot plasma. Their clouds are of similar column to those discussed here, but at higher density, and closer in. They find that soft excess and hard X-ray humps naturally go together.

(ii) Our clouds are closer and thicker than those deduced by \citet{Risaliti2007} and related papers to explain rapid absorption changes in many AGN. They may be more closely related to to the material producing Compton-thick partial covering phenomena revealed by BeppoSAX and Suzaku at $E>$10keV \citep{Dadina2004, Reeves2009, Turner2009a}, although it has been claimed that  such objects may also be explained with disc reflection \citep{Fabian2005}. From a study of 165 AGN with INTEGRAL data, \citet{Ricci2011} have proposed that the large reflection component seen in apparently mildly absorbed Type 2 AGN may be explained by Compton-thick clumps partially covering the source. At middling covering factors such opaque partial covering produces a striking spectral signature. At large, but still not complete, covering factors, the leaked-through spectrum may look quite normal, and if there is additional thinner absorbing material at larger radii, the spectrum may easily be mistaken for being Compton-thin when in fact it is largely Compton-thick. A consequence is that the observed X-ray luminosity may be systematically wrong. The effect on the luminosity function for obscured and unobscured AGN could produce an artificial dependence of obscuration on luminosity, as noted by \cite{Lawrence2010} and explored in more detail by Mayo and Lawrence (in preparation). 

(iii) The clouds we have discussed may also be related to the high velocity absorbers seen in highly ionised Fe \citep{Pounds2003, Chartas2003, Reeves2009}. These seem to be very common \citep{Tombesi2010}, and have typical velocities of 0.1-0.2c. It is generally assumed that they represent outflows, but there have also been claims of inflows at similar velocity \citep{Dadina2005, Tombesi2010}. If material is in rotational or turbulent motion at such velocities, this is quite consistent with the kind of blurring we have assumed in this paper. In at least one case \citep{Risaliti2005} the outflow velocity was seen to change substantially from one observation to another, suggestive of the kind of instability we have suggested for our UV-reflecting clouds. 
The columns deduced for the high velocity X-ray absorbers are large but somewhat less than we have been considering here, $\log{N_H}=23-24$ as opposed to $\log{N_H}>25$, but we note that partial covering will dilute the strength of the absorption lines. Our UV-reflecting clouds may also be related to the fast moving material involved in the proposed absorption explanation of the ubiqitous soft excesses seen in AGN \citep{Gierlinski2004, Middleton2007}

(iv) In section \ref{sec:timescales} I suggested that variations in covering factor could explain the UV variability. Likewise, covering factor changes could explain the observed X-ray variability, as argued by \citet{Abrassart2000}. Several papers have argued that absorption changes can explain the complex changes seen in the spectra of individual objects (e.g. \citealt{Miller2009} and references therein; see also the review by \citealt{Turner2009}). However, UV and X-ray variability cannot be explained by exactly the same reprocessing clouds. As discussed in section \ref{coordination}, X-rays and UV track each other well over medium-timescales, but X-rays also show extra short timescale variability. This is naturally explained if UV reprocessing is dominated by thick clouds at $\sim$30R$_S$, whereas X-rays are affected both by these clouds and by closer-in clouds as well. However, it is not obvious why X-ray variations sometimes seem to lag  the UV variations.

Overall, the clouds required to explain $N_H$ changes, X-ray variability and spectral changes, thick partial covering, soft excesses, and high velocity outflows are all likely to be somewhat related to our proposed UV-reflection clouds, but are not the same. A distribution of cloud properties is required, and it is not yet clear if a single model can reproduce everything consistently.

\subsection{Apparent UV source size}

Following equation 2 of \citet{Morgan2010}, the characteristic UV size of a thin accretion disc around a black hole with $M=10^9 M_\odot$ accreting at the Eddington limit with efficiency $\eta=0.1$ should be $\sim 10 R_S$. However we have found that reprocessing clouds could be producing a false UV continuum over a considerable radial scale, 10--100 $R_S$ - energy produced in the central region emerges on a larger scale. Potentially therefore reprocessing clouds can solve the size problem implied by gravitational microlensing studies (see section \ref{size}). Whether this works quantitatively is not yet clear - the effective size will depend on the density and covering factor of the clouds with radius, which obviously requires a physical model. Such a model of the physical distribution of the clouds is also needed to predict the chromatic behaviour of microlensing fluctuations.

\subsection{Next steps}

In this paper I have concentrated on the properties of clouds that could plausibly explain the worrying observed features of the UV bump and its variability, and demonstrating the plausibility of such clouds as an explanation. What is needed is a more detailed specific model. This is rather a challenge; to go beyond the plausibility demonstrated here one would need to specify the distribution of clouds in density, column, and radius, and to specify their velocity field. This is too large a volume of parameter space to blindly explore as a grid. It really needs a prediction from a physical model. However, there are some general features that could be explored first.

(i) There should be some testable dependence on black hole mass and accretion rate, as these properties together predict the hardness of the intrinsic continuum. As noted in section \ref{n12-clouds}, the relative strengths of Ly$\beta$ and HeII Ly$\alpha$, and hence the 1100\AA\ bump and the EUV peak, will depend on the temperature of the intrinsic continuum. This may produce measurable effects on the UV-FUV SED, and on the observed BLR line ratios. 

(ii) Another possible mass dependence comes from cloud size. The expected cloud size is $\sim 10^{13}$cm, but with a likely order of magnitude range. For quasar-size black holes, $M\sim 10^9 M_\odot$, this is a small fraction of the source size, so that any one cloud has a small covering factor, and any changes are global ones pertaining to the ensemble of clouds. For lower-luminosity Seyferts, with $M\sim 10^{6-7} M_\odot$, the movement and physical changes of individual clouds may be significant, producing qualitatively different changes.

(iii) Although I have argued that physically motivated cloud models are needed next, an obvious step is models with a small number of cloud zones, to see whether UV-reflection effects, X-ray reflection effects, X-ray absorption effects, and microlensing chromatic effects, can all be explained together without the various cloud populations destructively interfering as it were. This could also lead to insights which would help the design of simultaneous UV / X-ray monitoring campaigns.

\section{Conclusions}  \label{sec:conclude}

The problems with the accretion disc interpretation of the Big Blue Bump - the universal knee and FUV falloff, the ionisation paradox, the timescale issue, the co-ordination and amplitude-colour problems, and the microlensing size discrepancy - can all potentially be explained if EUV emission from the inner disc is reprocessed by dense clouds at a few tens times $R_S$. A variety of clouds may be present, but the ones that produce the desired effect have densities in the range $10^{11}<n<10^{13}$ cm$^{-3}$, and column densities $N_H>4\times10^{24}$. Such clouds are opaque and reflect most of the incident continuum, but recycle a substantial fraction of the EUV energy as emission lines, dominated by HeII Ly$\alpha$ and HI Ly$\beta$. Blurring this emission produces a localised false peak at 1100\AA , followed by a turn back upwards in the FUV. Producing such clouds requires only a small fraction of the mass available in the accretion disc, and does not require the accretion disc to be completely disrupted, and may be most likely to happen at a few  tens of $R_S$. Such clouds will feel a very strong outward radiation force, but they may be formed and destroyed on a faster timescale than they are ejected, so that there will not necessarily be a continous wind, but rather an unstably variable cloudy atmosphere. Variations in covering factor may explain the observed UV variability. 

Such clouds may be related to those deduced from various X-ray observations, but are not the same. There is likely to be a large range of cloud properties at various different radii in an AGN, with those in different parts of parameter space producing different effects. It remains to be seen whether a physically realistic model can explain all the observations simultaneously and consistently.

In the 1980s, there was confused argument over whether the ``blue bump'' was really due to an accretion disc as proposed by \citet{Malkan1982} -  some researchers claimed that the turn-up seen in optical spectra was actually due to Balmer continuum and FeII emission lines on top of an underlying power-law \citep{Grandi1982, Wills1985}. The solution was to take a wider (IR-optical-UV-X-ray) perspective and see that both phenomena were present. \citet{Elvis1985} first used the term ``little blue bump'' to distinguish the atomic feature from the ``big blue bump''. If it is correct that the 1100\AA\ peak is due to blurred atomic features, we may have to rename this feature the ``medium blue bump'', with the true ``big blue bump'' residing at 300\AA .


\section{Acknowledgements}  \label{sec:acknow}

In worrying about the problems with the big blue bump over a number of years,  I have talked to many people, but particularly influential conversations were had with Andrew King, Martin Elvis, Julian Krolik, Christine Done, Omer Blaes, and Ski Antonucci. Several very useful suggestions were also made by anonymous referee, particular concerning microlensing and covering factor issues. When I described how any dense clumps formed in the accretion disc would feel a very strong radiation force, Ski, colourful as ever, exclaimed that they would jump out of the disc ``like popcorn out of a pan''. 


\bibliographystyle{mn2e}
\setlength{\bibhang}{2.0em}
\setlength\labelwidth{0.0em}
\bibliography{lawrence-uvbump-arxivpaper-rev}

\label{lastpage}

\end{document}